\documentclass{amsart}
\usepackage{mathrsfs,amssymb,amsmath,amsfonts,amsxtra}
\usepackage{graphicx,color}
\begin{document}

\author{I. Schmelzer}
\thanks{Berlin, Germany}
\email{\href{mailto:ilja.schmelzer@gmail.com}{ilja.schmelzer@gmail.com}}%
\urladdr{\href{http://ilja-schmelzer.de}{ilja-schmelzer.de}}

\title{Pure quantum interpretations are not viable} \sloppypar

\maketitle
\begin{abstract}
Pure interpretations of quantum theory, which throw away the classical part of the Copenhagen interpretation without adding new structure to its quantum part, are not viable. This is a consequence of a non-uniqueness result for the canonical operators.
\end{abstract}

\sloppypar \sloppy
\newcommand{\B}{\mbox{$\mathbb{Z}_2$}} 
\newcommand{\Z}{\mbox{$\mathbb{Z}$}}
\newcommand{\N}{\mbox{$\mathbb{N}$}}
\newcommand{\R}{\mbox{$\mathbb{R}$}}
\newcommand{\C}{\mbox{$\mathbb{C}$}}
\newcommand{\pd}{\mbox{$\partial$}}
\newcommand{\p}{\mbox{$\hat{p}$}} 
\newcommand{\q}{\mbox{$\hat{q}$}} 
\newcommand{\ps}{\mbox{$\hat{p}(s)$}} 
\newcommand{\qs}{\mbox{$\hat{q}(s)$}} 
\newcommand{\h}{\mbox{$\hat{h}$}} 

\renewcommand{\H}{\mbox{$\mathcal{H}$}} 
\renewcommand{\L}{\mbox{$\mathcal{L}^2$}} 
\renewcommand{\O}{\mbox{$\hat{\mathcal{O}}$}} 
\newcommand{\V}{\mbox{$\mathcal{V}$}} 
\providecommand{\abs}[1]{\lvert#1\rvert}

\newcommand{\Sch}{Schr\"{o}dinger\/ }

\section{Introduction: The non-uniqueness of the canonical structure}

In \cite{kdv} we have proven two non-uniqueness theorems: For some fixed Hamilton operator \h, we have constructed for some continuous parameter $s$ different pairs \qs, \ps\/ of canonical operators so that
\begin{equation}\label{eH}
\h= \frac{1}{2m}\ps^2+V(\qs,s)
\end{equation}
with physically different, but equally nice (smooth, bounded, descreasing in infinity) potentials $V(q,s)$. In addition, we have constructed different tensor product structures (or ``decompositions into systems'') so that \h\/ has an equally nice, but physically different representation of type
\begin{equation}
\h=\sum\frac{1}{2m_i}\p_i(s)^2+V(\qs,s)
\end{equation}
in all or them.

From point of view of canonical quantization, there seems nothing problematic with this result. It nicely corresponds to the standard way to define canonical quantum theories: One has to define an irreducible representation of the canonical operators \p, \q\/ (with $[\p,\q]=-i\hbar$) and then to define the Hamilton operator \h\/ as a function of these operators
\begin{equation}\label{eCanonical}
\h= \frac{1}{2m}\p^2+V(\q).
\end{equation}
Once the theory is defined in such a way, no non-uniqueness problem appears -- the canonical operators \p, \q\/ are those used in the definition of the canonical theory.

The situation is different if we consider interpretations of quantum theory. The point is that the interpretation has to define the physical meaning not only for the Hamilton operator \h, but for all physically relevant parts of the theory. Once different choices of $s$, thus, identifications of \p, \q\/ with different \ps, \qs\/ would lead to physically different theories (with different potentials $V(q)$), the operators \p, \q\/ are physically important, thus, the interpretation has to describe their physical meaning too. In \cite{kdv} we have already considered the consequences of these non-uniqueness results for applications of decoherence in fundamental physics: The widely held belief that decoherence allows to define the classical limit without additional structure has to be given up. We have also evaluated and rejected the idea to postulate some fundamental decomposition into systems to derive a preferred basis. An emergent configuration space $Q$ would lead to a lot of losses (uncertainty, dependence on dynamics) which are in no way compensated by gains in explanatory power.

The aim of this paper is to continue the consideration of the consequences of these non-uniqueness theorems, in particular for various interpretations of quantum theory.

First, we discuss and reject the proposal to embrace the different \qs, \ps\/ as many different but equally real worlds -- an idea close to but not identical with ``many worlds''. Once this proposal is rejected, an interpretation has to identify the ``correct'' canonical operators \q, \p\/ among the \qs, \ps. We argue that this requires more than assigning pure labels. The canonical operators \p, \q\/ are in some sense different from the alternatives \ps, \qs, a difference which is not part of the mathematical formalism of pure quantum theory and has to be defined by the interpretation. This requires some additional physical structure which the interpretation has to define. 

Some interpretations have adequate structures -- pilot wave theories \cite{Bohm,deBroglie}, Nelsonian stochastics \cite{Nelson} and physical collapse theories \cite{GRW}, \cite{Ghirardi} have a preferred configuration space $Q$, and the Copenhagen interpretation associates the \q, \p\/ with classical measurement procedures. But there is a whole class of interpretations which does not have such a structure -- a class we name ``pure interpretations''. Such interpretations are the result of a very natural approach: One the one hand, one wants to get rid of the uncertain, problematic ``classical part'' of the Copenhagen interpretation. One the other hand, one does not want to introduce additional structure into the theory. A reasonable minimalistic approach, and it would be nice if it would work. But, as a consequence, the Copenhagen solution of our non-uniqueness problem no longer works, and without any new structure no new solution is available. Thus, this minimal, pure program fails.

We discuss shortly some important examples of such interpretations: Mermin's Ithaca interpretation \cite{MerminIthaca}, consistent histories \cite{Griffiths}, and the Everett interpretation \cite{Everett}. The aim is not to give a complete list of interpretations endangered by our non-uniqueness result. While I think that the problem is sufficiently general, so that every interpretation deserves a consideration if and how it solves this non-uniqueness problem, this paper can be only a starting point. Our examples merely illustrate that the problem appears in quite different approaches to the interpretation of quantum theory. 

If interpretations which have a non-uniqueness problem will be given up or saved by introducing some additional structure is a decision we of course have to leave to the proponents of these interpretations. Only they can be expected to find the optimal solution for their preferred interpretations. The point of this paper is to clarify some general, common points -- that, if one removes the classical part of the Copenhagen interpretation, one has to introduce something else, a replacement. This replacement has to be a non-trivial physical structure, which contains sufficient additional information to identify the canonical operators. The identification of the canonical operators with momentum and position measurements, which is postulated in the Copenhagen interpretation, has to be derived now based on the new physical structure.

This fact alone already removes one of the main advantages of the Everett program and similar programs. What can be obtained in this way is no longer a pure, minimal interpretation, but only an interpretation with some additional structure. The question is no longer if an interpretation has some additional structure (with automatic rejection of interpretations which have such ``hidden variables'') but what is the additional structure connected with a given interpretation, what are their particular advantages and disadvantages (now with automatic rejection of interpretations without such a structure).

The loss of purity is not the only possible consequence of the additional structure. Other questions may be influenced too. We consider two examples: First the popular many world argument ``[P]ilot-wave theories are parallel-universe theories in a state of chronic denial'' \cite{Deutsch} becomes invalid if (as one would expect) the additional structure introduced into many worlds is not part of pilot wave theory. On the other hand, if we introduce one ``preferred'' consistent family of histories as the additional structure into the consistent histories interpretation, this interpretation becomes more compatible with classical logic and realism, which would remove some arguments against it. But there are lots of other problems to be considered in future research. In particular, the symmetry properties of the theory may be heavily influenced by an additional structure.

\section{What is wrong with ``many laws''}\label{manyworlds}

Anticipating a possible non-uniqueness of the construction of a preferred basis, Saunders \cite{Saunders} has proposed a solution which  avoids the introduction of new physical structure: One could accept the non-uniqueness and consider all the different classical limits defined by different \ps, \qs\/ as equally real different worlds. Brown and Wallace \cite{BrownWallace} describe this idea in the following words:
\begin{quote}
Suppose that there were several such decompositions, each supporting information-processing systems. Then the fact that we observe one rather than another is a fact of purely local significance: we happen to be information-processing systems in one set of decoherent histories rather than another.
\cite{BrownWallace}
\end{quote}
This looks like a many-worlds-like solution of the problem. But it should be noted that there is an essential difference between usual many worlds and this proposal. The important point is that, as we have shown in \cite{kdv}, the different \ps, \qs\/ define different physics. They have even different classical limits $H(p,q,s)= \frac{1}{2m}p^2+V(q,s)$.  This situation is very different from the standard ``many worlds'', where all worlds follow the same physical laws, only with different initial values. We have not only ``many worlds'', but these many worlds follow different physical laws, a situation which is better named ``many laws''.

This difference allows some counterargumentation.

\subsection{Loss of predictive power}

The first one is based on Popper's criterion of empirical content. Its absolute version is that scientific, empirical theories should be falsifiable, they should allow the derivation of testable predictions. But we need the relative version, which allows to compare the empirical content of different theories. If theory $\mathcal{T}_1$ makes at least one prediction which cannot be made by theory $\mathcal{T}_2$, but all predictions of theory $\mathcal{T}_2$ are also predictions of theory $\mathcal{T}_1$, then theory $\mathcal{T}_1$ has more predictive power, or higher empirical content in comparison with $\mathcal{T}_12$. Popper's criterion tells us that the theories with higher empirical content should be preferred.

Let's apply this to our situation, and let's compare the empirical content of a single law theory (which has somehow fixed one value of $s$ and chosen the corresponding set of canonical operators \p, \q) with a many laws variant which embraces all the \ps, \qs\/ as different really existing worlds. We have shown in \cite{kdv} that the physical predictions for at least some experiments depend on $s$. The single law theory where $s$ is fixed allows to make the usual predictions of canonical quantum theory. The situation for the many laws version is different. This version cannot exclude that the value of $s$ in our universe is the same as that chosen by the single law theory, simply because that single law is one of the possible many laws. Therefore it is in principle impossible to falsify the many laws version without falsifying the single law version.

In the other direction this is possible: The single law theory may be falsified simply because the value of $s$ is wrong. Because different $s$ lead to different physical predictions, it may happen in principle that we observe an effect as predicted for a value $s'$ which is different from the $s$ used in the single world theory. This would falsify the single world theory. But if the correct value  $s'$ of our universe is among the allowed values in the many laws theory, many laws is not falsified by this observation. So many laws theory remains unfalsified -- the other, correct value $s'$ is allowed, it is simply the actual value of some other world.

Thus, the consideration of the predictive power gives a clear answer. A single world theory has more predictive power, higher empirical content, is able to make more specific predictions, and, therefore, has to be preferred following Popper's criterion. This is a natural consequence of the fact that the canonical operators \p, \q\/ of the theory are fixed and well-defined.

\subsection{A symmetry argument}

Let's add a completely different argument, which is based on symmetry. Once in the many laws version all worlds are equally real, have the same ontological status, the physical properties of our particular world cannot be further restricted by Ockham's razor or further symmetry principles. These principles allow to restrict only theories about what really exists. All the really existing worlds are already on equal footing.

In this case, one would expect that the law of our actual universe should be a typical, generic element of the set of laws. Indeed, if our observed law would be a very special element, say, for the sake of the argument, the one having a special value $s=0$, then there would be no point of considering all the other laws. One could simply use Ockham's razor to cut all the laws with $s\neq 0$ out of the theory.

Is the law of our actual universe a typical, general element of the set of all possible laws?  This is something we cannot decide, given that we have considered in \cite{kdv} essentially only the one-dimensional case (a two-dimensional construction was based on a product of the one-dimensional case), thus, don't know nor the real law of our universe, nor the possible modifications of it if we allow for other choices of the canonical operators. But let's nonetheless use the one-dimensional case considered in \cite{kdv}, with Hamiltonians of type \eqref{eH}, as a toy model of the possible laws for the universe, so that another choice of the canonical operators corresponds to another choice of $s$ in \eqref{eH}. What would be, in this toy-many-laws theory, the analogon of the law of our universe? Given the important role of the potential $V(q)=1/\abs{q}$, this potential seems to be the only reasonable candidate.

Can the potential $V(q)=1/\abs{q}$ be considered as a typical, generic element of some class of potentials $V(q,s)$ connected with each other by different choices of the canonical operators? The answer is a clear no. The potential $V(q)=1/\abs{q}$ can be characterized among them by an extreme symmetry property.

To see this symmetry property we have to ask what changes if we change $s$ in terms of the eigenstates and eigenvalues of the Hamilton operator. The answer is that the eigenvalues remain unchanged (the operator \h\/ remains the same by construction, as an abstract operator in the Hilbert space, and the eigenvalues $E_k$ are completely defined by the operator alone). The eigenstates $|\psi_k\rangle$ themself, as abstract elements of the Hilbert space, remain unchanged too. But their positions $q_k=\langle\psi_k| \qs |\psi_k\rangle$ change, because they depend on the operator \qs\/ which changes. This change is not a simple common shift -- this would be the result of a simple shift in the potential $V(q)\to V(q-q_0)$. Different eigenstates obtain different shifts, and explicit formulas for these shifts can be found from the theory of the KdV equation \cite{book-kdv}.

Now, the positions of the energy eigenstates for $V(q)=1/\abs{q}$ can we described in a very simple way: they are simply all zero. Thus, in our toy model our universe is distinguished by the very special symmetry property that $\forall k\; q_k=0$. This is certainly not a generic element. It would need high conspiracy. Thus, we have (at least for our toy model) a fine tuning problem: The ``many laws'' approach would lead to the expectation that the $q_k$ are a quite arbitrary sequence of real numbers. Instead, our choice of the potential leads to the very special case $q_k=0$ for all $k$.

For a theory with a single law this would be the most natural choice, clearly preferred by Ockham's razor. Instead, for a theory of may laws this is an extremal property of our own universe which requires explanation. 

\subsubsection{But what about our real world?}

A natural objection is that our one-dimensional toy consideration is much too trivial and our choice of the potential $V(q)=1/\abs{q}$ much too artificial to be relevant for our universe. So let's try to find out which part of the toy argument can be expected to generalize to a more general, high-dimensional, realistic situation. First, of course, it remains unchanged that the eigenstates $E_k$ of \h\/ remain unchanged. But once the eigenstates of \h\/ are unable to fix the position operator \q\/ and the potential $V(\q)$ already in the simplest one-dimensional case, one would not expect that this changes in higher dimensions (even if the nice exact mathematics of the Korteweg-de Vries equation works only in the one-dimensional case). A modification of \q\/ will also change the positions of the eigenstates $q_k=\langle\psi_k|\q|\psi_k\rangle$ of \h, since they obviously depend on \q. The question is how reasonable it is to assume some symmetries like $\forall k\; q_k=0$. But this is quite common already in the one-dimensional case. All we need in this case is the discrete symmetry $V(q)=V(-q)$. In this case, for an assumed asymmetric eigenstate $\psi_k(q)$ with $q_k\neq 0$ $\psi_k(-q)$ would be an eigenstate of the same energy $E_k$, and then their symmetric and antisymmetric combinations
$\psi_\pm(q)=\psi_k(q)\pm \psi_k(-q)$ would define other eigenstates already with $q_k=0$, so that every eigenstate can be represented as a linear combination of the same eigenvalue with $q_k=0$. (In the one-dimensional case this would be even easier, because in this case there are no degenerate eigenstates.) One can imagine that almost every symmetry which acts nontrivially on \q\/ (generalizing the $\q\to -\q$ of our example) may have similar nontrivial consequences for the positions of the eigenstates. Given the large role of symmetry in modern physics it seems quite reasonable to expect that there will be some symmetries in the final theory of everything too. Therefore the key elements of our toy example seem to have at least a chance to carry over to the situation of our real world.

\subsubsection{Maybe the symmetry helps?}

But maybe the symmetry we have mentioned here allows to solve the very problem? We prefer, of course, symmetric theories. And, of course, if we have to choose between a symmetric theory and one without symmetry we prefer the symmetric theory. But the very point of a ``many laws'' theory is that the theory itself does not make such a choice.  The ``symmetric theory'' is, in this concept, not a separate symmetric theory, but only a particular universe, with a particular law which has a particular symmetry. We do not have to make a choice between theories -- there is only one theory, which contains different worlds with different laws. So, Ockham's razor or human preferences for symmetric theories are of no use here. A special symmetry of the laws of our particular universe is something which requires explanation.

The situation becomes different if we reject the many laws proposal and want to use this symmetry to find a preferred set of canonical operators. For this purpose, symmetry properties of some choices of \p, \q\/ may be useful, and we will clearly prefer a more symmetric choice.

\subsubsection{Maybe anthropic argumentation helps?}

The laws of our world may be not a typical element in the set of all possible laws. Last but not least, our laws allow the existence of human beings. Maybe anthropic considerations allow to fix the parameters $s$ so that no conspiracy is needed?

Now, anthropic arguments seem quite weak in their ability to restrict parameters. The general picture, as defended, for example, by Weinberg \cite{Weinberg}, is that some parameters may be restricted in some regions of the parameter space by anthropic considerations, other parameters not. For some parameters may exist wide ranges where anthropic considerations are irrelevant, because these parameters do not seem to influence anything relevant for human existence. For example, the cosmological constant $\Lambda$ should be small enough to allow human existence. But if it is below a certain limit $\Lambda_0$, anthropic considerations are unable to tell anything. And if $\abs{\Lambda}\ll \Lambda_0$ the fine tuning problem is not solvable by anthropic considerations. For other parameters, anthropic considerations may not exclude anything interesting. For example, the mass of the top quark seems quite irrelevant for everything related with humans. It could influence something only if it would be many orders lower than it is. But nobody would observe any important difference if it would be many orders greater than it is. Thus, the predictive properties of anthropic considerations seem to be quite restricted.

Let's see what would be required. The construction as given in \cite{kdv} contains only one parameter $s$, but can be easily extended to an infinite set of parameters $\{s_{2k+1}\}$ where $s=s_3$ is related with the KdV equation itself, $s_1$ defines a simple shift, and the other $s_{2k+1}$ are related with other differential equations of order $2k+1$, so-called higher KdV equations (see, for example, \cite{book-kdv}). And for each of these parameters we have a similar situation: Different $s_{2k+1}$ define different physics, with a different potential $V(q,s_{2k+1})$, changing the operators \p, \q, but leaving \h\/ unchanged. Thus, the non-uniqueness problem is a multi-dimensional one, all the parameters $s_{2k+1}$ would have to be restricted. Similarly, if we consider, instead, the positions of the eigenstates $q_k$ as the parameters to be restricted to $q_k\approx 0$, we also have an infinite set of parameters.

So, even if one can reasonably hope that anthropic considerations may restrict a few of the $s_{2k+1}$, or some of the $q_k$, what would be the base for the hope that it allows to restrict all of them?

\subsection{Summary}

We have here even two independent arguments against a ``many laws'' proposal, of quite different character. A simple methodological one using Popper's criterion of empirical content and fine tuning argument based on the thesis that our laws are more symmetric than the average laws in this construction. If these two arguments are sufficient to convince proponents of this idea is another question. Given that the proposal has been made by Saunders \cite{Saunders} in a situation where the non-uniqueness construction of \cite{kdv} was yet unknown, and that Tegmark has proposed an even more radical version of Platonism where every imaginable mathematical universe really exists \cite{TegmarkMath}, the idea to extend many worlds ruled by a common law into many laws seems to be attractive to many worlders on its own right, even without the necessity to solve the non-uniqueness problem.

But in the remaining part of this paper we will ignore the ``many laws'' possibility. So in the following we assume that there is only a single law of physics, which makes certain predictions. Thus, it follows from the experiment considered in \cite{kdv} (which proves that the different choices lead to different physical predictions) that the complete description of physics consists not of \h\/ alone but also of the canonical operators \p, \q. The alternative choices \ps, \qs\/ are unphysical.

\section{Pure interpretations: The minimal program for replacement of Copenhagen}

So, assume that we don't want to embrace all \ps, \qs\/ as describing different real worlds. There is only one \p, \q, which describes the true momentum and position measurements, while all the other  \ps, \qs\/ are only mathematical constructions without any physical meaning. They may describe some other measurements, but nobody knows which, and nobody cares.

But the \p, \q\/ are not distinguished among the \ps, \qs\/ by their mathematical properties. Each defines an irreducible representation of the canonical commutation relations: $[\ps,\qs]=-i\hbar$. Each of them is connected with the Hamilton operator \h\/ in a quite similar way -- the Hamilton operator has the same form \eqref{eH}, and the potential $V(q,s)$ has similar nice qualitative properties. But something in the interpretation should tell us why we nonetheless have to use the operators \p, \q\/ (instead of one of the \ps, \qs) if we want to measure the momentum or the position.

In the Copenhagen interpretation this is done. We have the canonical operators \p, \q. And these canonical operators are defined as describing the momentum and position measurements. What do these phrases ``momentum measurement'' and ``position measurement'' mean? The answer is contained in the classical part of the Copenhagen interpretation. Or at least supposed to be. In fact, this classical part is not formalized, and there seems to be no hope that such a notion as ``momentum measurement'', which covers lots of very different macroscopic measurement devices, can be really made precise and certain.

That's why this ``classial part'' of the Copenhagen interpretation has been widely considered as unsatisfactory, and has motivated attempts to get rid of it. The ideal solution would be a derivation of the classical part from the quantum part taken alone. The program to find an interpretation of this ideal type, which reject the classical part of Copenhagen and start with the pure quantum part, without introduction of additional structure, we call \emph{pure program}, and the resulting interpretations (even if they are in fact not completely realized) \emph{pure interpretations}.

The most popular example is the Everett interpretation (better named ``Everett program''), described by Everett in the following way:
\begin{quote}
``This paper proposes to regard pure wave mechanics \dots as a complete theory. It postulates that a wave function that obeys a linear wave equation everywhere and at all times supplies a complete mathematical model for every isolated physical system without exception. \ldots The wave function is taken as the basic physical entity with no a priori interpretation. Interpretation only comes after an investigation of the logical structure of the theory. Here as always the theory itself sets the framework for its interpretation. \ldots
The new theory is not based on any radical departure from the conventional one. The special postulates in the old theory which deal with observation are omitted in the new theory. The altered theory thereby acquires a new character. It has to be analyzed in and for itself before any identification becomes possible between the quantities of the theory and the properties of the world of experience.'' \cite{Everett}
\end{quote}
There are, of course, lots of variants of many worlds interpretations, and not all of them follow the original pure program. But our point is, of course, not to criticize non-pure variants of many worlds: Instead, our point is that the original, pure program is not viable.

Many worlds is not the only such program. Another example is Mermin's ``Ithaca interpretation'' (also more appropriately named ``Ithaca program'') \cite{MerminIthaca}: On the one hand, Mermin tells that ``\ldots I would like to have a quantum mechanics that does not require the existence of a classical domain'' and introduces as one of the desiderata ``The concept of measurement should play no fundamental role''. On the other hand, we read that ``\ldots by quantum mechanics I mean quantum mechanics as it is -- not some other theory in which the time evolution is modified by non-linear or stochastic terms, nor even the old theory augmented with some new physical entities (like Bohmian particles) which supplement the conventional formalism without altering any of its observable predictions.''

Thus, the basic idea of a pure quantum interpretation is shared by quite different approaches. But this is quite natural. The most questionable part of the Copenhagen interpretation is its classical part, so it is natural that one wants to get rid of it. On the other hand, one wants to minimize the number of assumptions one has to make. And the minimum would be reached if nothing would be added.

Unfortunately, because these pure interpretations reject the classical domain of the Copenhagen interpretation, the Copenhagen way to solve our non-uniqueness problem has been lost. The ``correct'' operators \p\/ and \q\/ can no longer be distinguished among the \ps, \qs\/ by a postulated association with specific experimental arrangements described in classical language -- this is part of what has been removed. On the other hand, once pure interpretations refuse to add some replacement, some new, additional structure, they seem unable to compensate for the loss.

Thus, the non-uniqueness result of \cite{kdv} shows that the ``pure program'' -- the program to develop pure interpretations of quantum theories -- cannot be realized and has to be given up. If one removes the association of the canonical operators \p, \q\/ with certain measurement procedures, which is defined by the classical part of the Copenhagen interpretation, one has to add something else, something which allows to identify the \p, \q, in some other way with momentum and position measurements we make. For this, the \p, \q\/ have to have some special properties which distinguish them from all the other \ps, \qs.

\section{``Special'' interpretations?}

Let's return now to the hope that some special symmetries of the problem may be used to distinguish the  \p, \q, by their special properties from the other  \ps, \qs. In our toy example this has been the property that all the positions of the eigenstates $q_k=\langle\psi_k|\q|\psi_k\rangle$ are zero.

Let's note here an important and interesting difference between an imagined interpretation based on such an idea and existing interpretations. Canonical quantization works for arbitrary potentials $V(q)$, and the Copenhagen interpretation does not object and gives all these canonical quantum theories the same sort of interpretation. But we can do canonical quantization for all the potentials $V(q,s)$, and the theory defined by \h\/ as given by \eqref{eH}, and \p\/ and \q\/ as given by \ps\/ and \qs\/ defines a realization of the canonical quantization for the potential $V(q,s)$. If an interpretation forgets now about the canonical operators \p, \q\/ as defined by the canonical quantization procedure, and makes a new choice of the \p, \q\/ based some mathematical properties of their relation to \h\/ (like the property $q_k=0$) then we do not recover in the classical limit the original theory we have canonically quantized -- with potential $V(q,s)$ -- but another one, with the potential $V(q,s')$ for the preferred $s'$.

Thus, if we follow this strategy, we obtain a lot of changes in the general picture of quantization: Only a few potentials, distinguished by some special properties, allow to be quantized.  Canonical quantization of other potentials gives, of course, a canonical quantum theory as before, but the classical limit of this theory, as described by an interpretation of this type, leads to a different classical theory with different potential. For example, if we would use the property $\forall k\; q_k=0$, the corresponding potentials would have the symmetry $V(q)=V(-q)$, and only potentials of this type could be obtained as a classical limit of quantum theories in this interpretation.

This is not obviously false -- essentially, to be viable, an interpretation should be able to quantize only a single potential, the one we observe in our universe. Then, such a restriction of the allowed potentials is a testable, falsifiable prediction, also something nice. And if the potentials preferred by this interpretation have, moreover, nice additional symmetry properties, as the $V(q)=V(-q)$ symmetry of our toy example, this gives the interpretation some advantage in beauty.

So this may be an interesting way to meet the non-uniqueness problem of \cite{kdv}. But it does not meet the criteria of the ``pure program'', because it adds new physics, even important new physics, by restricting the class of potentials $V(q)$ allowed in canonical quantum theories.  Thus, an interpretation of this type is physically very different from the Copenhagen interpretation (which does not make any such restrictions). And, in particular, if we start with a canonical theory which has a ``wrong'' potential, an interpretation of this type makes physical predictions different from the Copenhagen interpretation.

Thus, the thesis in the title of our paper is not endangered by interpretations of this type. Therefore we can ignore them in the remaining part of the paper and focus our interest on interpretations which are not ``special'' in this sense, interpretations which handle all potentials $V(q)$ on equal foot.

\section{The necessity of a new physical structure}

That means, we assume that they allow canonical quantization for all sufficiently well-behaved potentials $V(q)$, and position and momentum measurements, as described on the base of these interpretations, have some association with the canonical operators. Note that in interpretations which do not contain the Copenhagen classical part this association has to be derived, because it is no longer postulated. And we assume that this derivation, whatever the potential $V(q)$ used in the theory, recovers at least approximately the Copenhagen association of the canonical operators with the canonical measurements and recovers also the classical limit.

The point we want to make in this section is that this requires the introduction of some additional physical structure.

\subsection{Pure labels are not sufficient}

The problem is that a canonical quantum theory in itself gives only labels. There is some operator denoted \p\/ with the label ``momentum operator'', some other operator denoted \q\/ with the label ``position operator'', which form an irreducible representation of their commutation relations $[\p,\q]=-i\hbar$. The Hamilton operator \h\/ has the form \eqref{eCanonical} with some potential $V(q)$. That's all what is given by the canonical quantum theory itself.

Now let's compare this with some other choice of $s$. We have now another operator, denoted here \p', which has now the label ``momentum operator'', and also another operator \q' which has the label ``position operator''. But in all other aspects this is a standard canonical quantum theory. Thus, there is some (other) potential $V'(q)$, but the general form of the Hamilton operator \h\/ in terms of the \p', \q' is the same canonical form \eqref{eCanonical}. This other theory is simply equivalent to standard canonical quantum theory for a different potential $V'(q)$.

But as a consequence of our assumptions, momentum and position measurements for these two theories have to be different. In the first canonical theory, we obtain the predictions for momentum and position for the potential $V(q)$, in the second theory those for the potential $V'(q)$.  But the canonical theories themself have distinguished the operators \p\/ and \q\/ only by giving them different labels. The operator \h\/ which defines the time evolution was the same, the operators \p\/ and \q\/ define an unitarily equivalent representation of the same commutation relation $[\p,\q]=[\p',\q']=-i\hbar$. Pure labels don't change anything. Thus, no physical predictions should differ simply because we have decided to name \p' instead of \p\/ the ``momentum operator''.

Thus, there should be also something else which changes together with the label ``momentum operator''. Something physical, because it leads to differences in the physical predictions, in particular for momentum measurements. There should be some connection between the label ``momentum operator'' and physics, a connection which is not covered by the mathematics of canonical quantum theories (that means, by the irreducible representation of the canonical commutation relations and the general form \eqref{eCanonical} of the Hamilton operator), but which allows to identify the expectation value $\langle\psi|\p|\psi\rangle$ of the operator \p\/ (the one with the label ``momentum operator'') with the expectation value of some real physical experiment which measures momentum.

In the Copenhagen interpretation, such a connection exists -- the association between the label ``momentum operator'' and the momentum measurement is simply postulated in the classical part of this interpretation.  Removing this part of the Copenhagen interpretation would remove this association, reducing ``momentum operator'' to a pure label without association to measurement procedures. But we have to recover this association, because this is what the theory predicts. So there should be something else, some physics defined by the interpretation, which allows to establish such an association.

To clarify what is meant with a new physical structure, let's consider a few examples. 

\subsection{Theories of pilot wave type}

With ``theories of pilot wave type'' we mean a large number of quite different interpretations. First, there are of course de Broglie-Bohm pilot wave theories \cite{Bohm,deBroglie} with different choices of the configuration space. But we include here also some stochastic theories like Bell-type field theories \cite{BellFT} and Nelsonian stochastics \cite{Nelson}. To combine them all into a single class is justified only because for the question discussed here their differences do not matter. They use, essentially, the same type of additional physical structure -- an explicit trajectory $q(t)\in Q$ in the configuration space $Q$. This trajectory may be deterministic in pilot wave theories theories, stochastic in Nelsonian stochastics, and even discontinuous stochastic in the case of discrete configuration spaces $Q$. But in all these theories we have a new physical law, a variant of the ``guiding equation'' of pilot wave theory, which defines the evolution of the configuration $q(t)$.

Then it is postulated what our own state is described by the configuration $q(t)$ instead of the wave function $\psi(q)$. This postulate allows to identify measurements as something which has to change the physical state of our brain, thus, as something which changes the value of some particular configuration variables $q_{brain}(t)$. This is a sufficient base for the development of a measurement theory. In particular, in the classical limit we obtain the classical trajectory $q(t)$ simply as the limit of the quantum trajectory $q(t)$.

The new physical structure, therefore, consists of the following elements: The identification of the configuration space $Q$, or, in other words, of the operator \q, a new equation for the evolution of $q(t)$, and the identification of the state of the universe with the configuration of the universe $q$. The canonical operator \q\/ has therefore a direct connection with the new structure, which does not exist for the other \qs. The canonical theory based on the other operator \q' leads to a very different pilot wave theory, with another configuration space $Q'$. The trajectory $q(t)$ of the first theory and the trajectory $q'(t)$ of the second theory have nothing to do with each other -- we cannot even compare them because they live in different spaces.

\subsection{Physical collapse theories}

Another class of interpretations which have introduced a new physical structure are physical collapse theories \cite{GRW,Ghirardi}. In these theories the new physics consists of a modification of the \Sch equation. Some additional physical collapse mechanism disturbs the unitary \Sch evolution and leads to a localization of the wave function. This localization happens in the position representation $\psi(q)\in\mathcal{L}^2(Q,\C)$.

Given the collapse mechanism, we do not have to consider all wave functions $\psi(q)$ in the classical limit, but only a small subclass of localized wave functions $\psi(q,t) \approx \delta(q-q(t))$ which are localized around the classical trajectory.

The new physical structure in these theories is defined by the terms which modify the \Sch equation. These terms depend on something which explicitly depends on the canonical operator \q. Thus, the wave functions obtained in different canonical quantum theories follow different equations, and the classical trajectories $q(t)\in Q$ and $q'(t)\in Q'$ approximating them have nothing to do with each other.

The author prefers theories of pilot wave type, considering de Broglie's old argument that ``[I]t seems a little paradoxical to construct a configuration space with the coordinates of points which do not exist'' \cite{deBroglie} as sufficiently strong. But this preference is irrelevant for the question considered in this paper. What is interesting here is that above classes of theories have a sufficient additional physical structure and therefore no non-uniqueness problem.

\subsection{Predefined subsystems}

In the previous examples, the configuration space $Q$ itself has already played a special physical role. But in principle the new structure may be of a quite different type. A nice example to illustrate this is a predefined subdivision $\mathcal{H}=\mathcal{H}_A\otimes\mathcal{H}_B$ of the Hilbert space into physically different subspaces, for example into a bosonic and a fermionic part, as considered, for example, by Kent \cite{Kent}. Then we can apply techniques like  decoherence or the Schmidt decomposition to derive a preferred basis in one of them, defined, say, by some operator $\q_A$ on $\mathcal{H}_A$.

Here, the additional structure is defined by the subdivision $\mathcal{H}=\mathcal{H}_A\otimes\mathcal{H}_B$ itself and the connection between the Hamilton operator and this subdivision. In the simplest case, one could, for example, imagine a Hamilton operator
\begin{equation}
 \h = \frac{1}{2m_A}\p_A^2 + \frac{1}{2m_B}\p_B^2 + V(\q_A,\q_B)
\end{equation} 
which restricts observations of $A$ made by the $B$-part to measurements of $\q_A$.

In \cite{kdv} we have presented some arguments against interpretations of this type. But these arguments are irrelevant for the point of this paper, which is that we need an additional physical structure.  This additional physical structure is present, and it is also sufficient to identify the canonical operator, even if it is only the operator $\q_A$ of some part of the theory $\mathcal{H}_A$.

\subsection{Summary: What we need}

Thus, to solve the non-uniqueness problem, we need an additional physical structure.  The pure labels ``position operator'' and ``momentum operator'' which remain if we remove the classical part of the Copenhagen interpretation (which have given them an explicit, even if only postulated, connection with position and momentum measurements) are not sufficient. We need something which gives different predictions for these measurements for different choices of the canonical operators among the \ps, \qs.

This requirement is not too strong, as we have seen in some examples of interpretations which have such additional structures. To solve the non-uniqueness problem, it is important that, first, we have an additional structure, and, second, that this additional structure is sufficient to make a choice among the candidates for the canonical operators \ps, \qs.

\section{Consistent histories}

And interesting example where the question if the additional structure is sufficient is problematic is the ``consistent histories interpretation''. An implicit reference to it we have already cited -- our quote from  Brown and Wallace \cite{BrownWallace} which describes what we have named the ``many laws'' solution has used the language of consistent histories. And this seems not to be an accident. It seems quite reasonable to expect that in consistent histories  different choices of $s$ will be associated with different consistent families of histories..

In its intentions, the ``consistent histories interpretation'' seems close to a pure interpretation, despite the fact that it includes some additional structure. In particular, it rejects the classical Copenhagen part (``The interpretive scheme which results is applicable to closed (isolated) quantum systems, \ldots has no need for wave function `collapse,' makes no reference to processes of measurement (though it can be used to analyze such processes) \ldots'' \cite{Griffiths}). What it adds to the quantum formalism is ``\ldots an extension of the standard transition probability formula of nonrelativistic quantum mechanics to certain situations, we call them `consistent histories,' in which it is possible to assign joint probability distributions to events occurring at different times in a closed system without assuming that the corresponding quantum operators commute.''

Is the additional structure added by consistent histories sufficient to identify the canonical operators? The answer seems to be negative. Or, more accurate, I see no reason for hope that the consistency condition for families alone allows to distinguish between different standard quantum theories having the same standard form
\begin{equation}\label{eHsimple}
\h= \frac{1}{2m}\p^2+V(\q)
\end{equation}
only with different potentials $V(\q,s)$. But this is what would have to happen if the only thing added -- the consistency condition for families -- would allow to distinguish between the different $s$.

Indeed, let's look what we would need. We have a definition of histories, a definition of families of histories, and some consistency conditions for these families. What would solve the problem would be the identification of a history which has a natural association with the canonical operators \p, \q. Now, histories are by definition given only at some discrete times $t_i$, and at each moment of time the operators $E^\alpha(t_i)$ defining the possible events have to commute. But these seem to be purely technical complications. If there would be a consistent family of histories
\begin{equation}
\mathcal{H} = \{O_{jk}(t_0),O_{jk}(t_1),\ldots,O_{jk}(t_n)\}
\end{equation}
where each event $O_{jk}$ gives $\langle\p\rangle \approx p_j$, $\langle\q\rangle \approx q_k$ with appropriate accuracy $\Delta p$, $\Delta q$, one could consider this part of the problem as solved.

But the problem we want to solve is not simply that for one of the \ps, \qs\/ such a consistent family should exist. The problem is that the structure added by the consistent histories approximation should identify a single $s$. It would not be a very big problem if this is only an approximate identification modulo some $\Delta s$ such that measurement results of the operators \qs\/ cannot be distinguished from each other. But some preferred $s$ should be distinguished at least approximately.

What does it mean? If we forget a moment about \h, then all the canonical pairs \ps, \qs\/ are unitarily equivalent (that's how they have been constructed in \cite{kdv}). Thus, if there is a family of histories $\mathcal{H}$ associated with one \p, \q, we can simply use this equivalence to construct families of histories $\mathcal{H}_s$ associated with every \ps, \qs. The original $\mathcal{H}$ is consistent for the Hamilton operator \h. The question is if the $\mathcal{H}_s$  are consistent. In principle, they may not -- the question if $\mathcal{H}_s$ is consistent given the Hamilton operator \h\/ is unitarily equivalent to the question if  the original family $\mathcal{H}$ remains consistent if $V(q)$ will be replaced by $V(q,s)$.

Of course, the consistency of a family of histories depends on the Hamilton operator. So, in principle, it may be possible that it among the $V(q,s)$ there is only one value of $s$ such that the family $\mathcal{H}$ is consistent. But would you bet that this will happen? I will certainly not. The $V(q,s)$ are solutions of the Korteweg-de Vries equation with $s$ as the ``time'' parameter \cite{kdv}. Their minima are localized at very different positions $q$, but look qualitatively equally nice and have equally nice formal properties (they are smooth, decrease at infinity) and even some important properties like the eigenvalues of \h\/ are the same. What could be a base for the hope that a pure consistency requirement allows to prefer, among them, a single value of $s$? I cannot see anything supporting such a hope.

So, if the additional structure which the consistent histories interpretation adds to pure quantum theory is some variant of a consistency condition for families of histories, there seems no reason to hope that the non-uniqueness problem may be solved.

But is it possible to save consistent histories by adding more structure? This seems not only possible, there is even a natural candidate -- some preferred consistent family of histories. The various criteria for consistency of families may have different solutions, different families of histories which are consistent, nice and beatiful according to various criteria. But the different families are incompatible with each other. If we hear different incompatible histories in everyday life, we believe at most one of them, and even if we have not yet decided which story we believe, we do not doubt that at most one of the stories can possibly be true. So all we have to do is to apply this rule of common sense to consistent histories. That means there is one consistent family which is correct, and everything incompatible with this family of histories has to be rejected. Now, if this preferred family is somehow associated with one set of canonical operators \p, \q\/ but not with others, then everything is fine. This hypothesis seems already quite natural.

Thus, while consistent histories as it is (with various consistency conditions, but no definite choice of a preferred family of histories) seems unable to solve the non-uniqueness problem, it probably may be solved by introducing a preferred family of consistent histories.

\section{Consequences of the loss of purity}

Even if one accepts that one needs additional structure to solve the non-uniqueness problem, one may decide that particular attempts to develop pure interpretations have their own value and should not be given up, even if the initial hope to obtain a pure interpretation cannot be realized. This seems sociologically plausible in particular for the Everett program. 

What would be the consequences?  Most importantly, the previously pure interpretation would lose its most attractive property -- its purity.

\subsection{The fate of the denial-argument}

An example of an application of purity is the argument that ``[P]ilot-wave theories are parallel-universe theories in a state of chronic denial'' \cite{Deutsch}. This argument is quite popular in the many worlds community \cite{Zeh, Brown, Wallace}. Given the counterargumentation by Valentini \cite{ValentiniNoDenial}, one would concede far too much if one would accept this argument as somehow endangering pilot wave theory, even if it remains popular among the many worlders \cite{BrownDenial}.

But let's nonetheless assume, simply for the sake of the argument, that the argument in its original version is not completely invalid. Assume now that, to meet our non-uniqueness argument, one introduces some additional structure into many worlds. In this case, it is very probable that the argument becomes invalid, for the simple reason that the new many worlds structure is not part of pilot wave theory.  Pilot wave theory already has an additional structure -- the trajectory of the configuration $q(t)\in Q$ guided by the guiding equation -- which distinguishs a certain \q, thus, there is no reason to introduce anything else. But the argument works only if all real, physical parts of the many worlds interpretation are also real, physical parts of pilot wave theory and follow the same equations (like the \Sch equation for the wave function).

Thus, the argument depends on the purity of the many worlds interpretation relative to pilot wave theory: Everything which is physical in the Everett interpretation should be physical in pilot wave theory too. If we save many worlds by the introduction of some additional structure, the argument has to be reconsidered. If the new structure is not part of pilot wave theory too, the argument becomes invalid.

Of course, one cannot exclude completely that the many worlders use an additional structure which is also part of pilot wave theory, so that the denial-argument survives. But this would be a rather strange choice. Last but not least, the additional structure is the configuration $q(t)\in Q$. This hidden variable is considered as unnecessary today and does not seem to be the first candidate to be embraced by the many worlds community. Even more, one could even question if a theory which gives a $q(t)\in Q$ the status of reality is yet correctly classified as ``many worlds'' -- it may be better characterized as a variant of pilot wave theory. It would be much closer to the current many worlds approach if, instead, some fundamental ``decomposition into systems'' would be used as the additional structure. But once no such ``decomposition into systems'' is part of pilot wave theory, the denial argument would be dead in this case.

But this may be not the only loss. One of the major arguments in favour of many worlds as well as of other pure interpretations is their claimed compatibility with relativistic symmetry.
\footnote{
Given that a preferred frame allows a simple explanation of the SM fermions and gauge fields in terms of a condensed matter model \cite{clm}, relativistic symmetry does no longer seem to have the fundamental importance which is attributed to it by the relativistic tradition.
}
But whatever the additional structure, it may restrict the symmetry group of the theory. In particular, the configuration space itself -- the structure defined by the operator \q\/ we have to choose among the \qs\/ -- is (at least in its usual form) not covariant. The danger that some additional structure will destroy relativistic symmetry is recognized, for example, by Wallace, who notes that ``\ldots there seems to be no relativistically covariant way to define a world \ldots'' \cite{Wallace}.

\subsection{Does consistent histories have to modify logic?}

On the other hand, the loss of purity may also lead to improvements. The additional structure may destroy arguments against the interpretation. Here, our proposal to introduce a preferred family of histories into the consistent histories interpretation may be an example. This is not only a possibility to solve the non-uniqueness problem, but removes also another argument against consistent histories -- that it modifies classical logic without necessity.

The problem (discussed for example in \cite{Exchange}) is the incompatibility of different, separately consistent, families of histories. A situation where we have different sets of statements which are internally consistent but incompatible with each other is quite common in everyday life -- these are simply different incompatible theories. Because they are incompatible with each other, at most one of them can be true. And even if we do not have sufficient information to identify the true theory, if all of them, taken separately, are compatible with all we know, it would not change our certainty that at most one of the theories may be true. This is, essentially, the whole point of logic.

But in consistent histories this everyday situation is interpreted in a completely different way: ``Note that incompatibility, the fact that the two families cannot be combined, does not mean that one is `wrong' and the other is `right.' Seeking some law of nature which `chooses' one rather than the other is to misunderstand the nature of quantum descriptions.'' \cite{Griffiths2}. In other words, we have different incompatible theories, but they are somehow all equally true.

The consistent historians are, of course, aware of the straightforward logical consequence -- if two theories which are incompatible with each other are above true, one can construct contradictions. They have ``solved'' this problem by introducing a new rule of logic -- that one is not allowed to combine statements which belong to different families: ``Since both $\mathcal{F}_a$ and $\mathcal{F}_b$ are consistent families, the conclusions of a probabilistic analysis applied using just one of them while disregarding the other will be correct. However, the families are incompatible, and so these conclusions cannot be combined. One cannot say that at time $t_2$ the particle is both in a superposition state $c$ AND that it is moving on the upper trajectory $u$, for that would be meaningless in the same way that `$S_x = +1/2 \text{ AND } S_z = +1/2$' makes no sense.'' \cite{Griffiths2}.

This ``solution'' is not accepted as satisfactoy by the critics. It looks like there are true statements $A$ and $B$, but to combine them into the statement $A \text{ AND } B$, something which is always allowed in classical logic, appears to be forbidden. Is this some sort of modification of the rules of logic, some variant of ``quantum logic''? If yes, then it seems extremely difficult to justify it. Even if the very foundations of the scientific method, including the logic, are in principle open to discussion and modification, the justification for a modification of logic should be extremely strong. And, given that there are viable alternatives which do not require modifications of logic, this does not seem to be the case. If classical logic is not changed, then what is modified if the incompatibility of two families of statements does not mean that at most one of them can be right?

Whatever, all these problems with incompatibility simply vanish if we introduce, as an additional physical structure, one consistent family of histories -- the one which contains the histories which are really possible. Thus, the additional structure which seems necessary to solve the non-uniqueness problem would, by the way, solve also another weak point of the consistent histories interpretation. 

\section{Discussion}

We have argued that the only way to handle our non-uniqueness result is to make a choice among the \ps, \qs, and to associate the preferred \p, \q\/ with some physical structure powerful enough to distinguish them as associated with momentum and position measurements. This destroys the minimal program for the improvement of the Copenhagen interpretation -- to throw away the classical part of the Copenhagen interpretation (which solves this non-uniqueness problem) without adding any new physical structure to its quantum part. This ``pure program'' is unable to give viable interpretations, interpretations which are able to solve our non-uniqueness problem.  A quantum interpretation which does not embrace the classical part of the Copenhagen interpretation needs an additional physical structure.

It was not the aim of the paper to present a complete overview over all the proposed interpretations of quantum theory whose viability is endangered in the light of our non-uniqueness problem.  We have presented some examples -- the Everett, Ithaca, and consistent histories interpretations. These examples illustrate that the problem is relevant for quite different approaches to the foundations of quantum theory.

But it was also not the aim of this paper to claim that the particular interpretations we have considered cannot be saved. Instead, one can save them by adding some new physical structure. Sometimes reasonable candidates are already part of the mathematical apparatus, and giving them physical significance could even improve them. We have argued that this in the case of consistent histories if one introduces one consistent family of histories as preferred.

On the other hand, the introduction of an additional physical structure means also some sort of loss. There is clearly a loss of purity. The interpretation becomes in some sense more complicate. It possibly decreases the symmetry of the interpretation and destroys some arguments in favour of it. We have discussed this for the case of the ``pilot wave theory is many worlds in a state of denial'' argument.

The decision if the particular interpretations are worth to be saved, and what are the best ways to save them, the optimal choices for the additional structure, is something we leave to the proponents of these interpretations.

The interpretations preferred by the author -- theories of pilot wave type which have a physically preferred configuration space $Q$ with a trajectory $q(t) \in Q$ -- do not have any non-uniqueness problem. This means not only that the problem is solvable and solved by other interpretations. It means also that what was a strong argument against pilot wave theory -- the existence of such an additional structure -- becomes now a strong argument in its favour.


\begin{thebibliography}{99}

\bibitem{kdv} Schmelzer, I.: Why the Hamilton operator alone is not enough, Found. Phys. vol. 39, p. 486-498 (2009), \href{http://arxiv.org/abs/arXiv:0901.3262}{arXiv:0901.3262}

\bibitem{book-kdv} Ablowitz, M. J., Clarkson, P. A.: Solitons, nonlinear evolution equations and inverse scattering, London Mathematical Society Lecture Note Series, 149, Cambridge University Press, Cambridge (1991)

\bibitem{BellFT} J.S. Bell, Beables for quantum field theory, Phys. Rep. 137, 49-54, 1986


\bibitem{Bohm} Bohm, D: A suggested interpretation of the quantum theory in terms of ``hidden'' variables, Phys. Rev. 85, 166-193 (1952)

\bibitem{deBroglie} de Broglie, L., La nouvelle dynamique des quanta, in Electrons et Photons: Rapports et Discussions du Cinquieme Conseil de Physique, ed. J. Bordet, Gauthier-Villars, Paris, 105-132 (1928), English translation in: Bacciagaluppi, G., Valentini, A.: “Quantum Theory at the Crossroads: Reconsidering the 1927 Solvay Conference”, Cambridge University Press, and \href{http://arxiv.org/abs/arXiv:quant-ph/0609184}{arXiv:quant-ph/0609184} (2006)

\bibitem{Brown} Brown, H.R.,  Comment on Valentini, ``De Broglie-Bohm Pilot-Wave Theory: Many Worlds in Denial?'', \href{http://arxiv.org/abs/arXiv:0901.1278}{arXiv:0901.1278}

\bibitem{BrownWallace} Brown, H.R., Wallace, D.: Solving the measurement problem: de Broglie-Bohm loses out to Everett,  Foundations of Physics, Vol. 35, No. 4, 517 (2005) \href{http://arxiv.org/abs/arXiv:quant-ph/0403094}{arXiv:quant-ph/0403094}

\bibitem{Deutsch} D. Deutsch, Comment on Lockwood. British Journal for the Philosophy of Science 47, 222–228 (1996)

\bibitem{Everett} Everett, H.: "Relative State" Formulation of Quantum Mechanics, Rev. Mod. Phys. vol. 29. n. 3 (1957)



\bibitem{Ghirardi} Ghirardi, G. C. (2002). Collapse theories. In the Stanford Encyclopedia of Philosophy (Summer 2002 edition), Edward N. Zalta (ed.), available online at\hfill \phantom{.} \href{http://plato.stanford.edu/archives/spr2002/entries/qm-collapse}{http://plato.stanford.edu/archives/spr2002/entries/qm-collapse}

\bibitem{GRW} Ghirardi, G., A. Rimini, and T. Weber: Unified Dynamics for Micro and Macro Systems. Physical Review D 34, 470-491 (1986)

\bibitem{Griffiths} Griffiths, R. B.: Consistent Histories and the Interpretation of Quantum Mechanics, Journal of Statistical Physics, vol. 36, nr. 1/2, 219-272 (1984)

\bibitem{Griffiths2} Griffiths, R. B.: Quantum mechanics without measurements, \href{http://arxiv.org/abs/arXiv:quant-ph/0612065}{arXiv:quant-ph/0612065} (2006)



\bibitem{Kent} Kent, A., Real World Interpretations of Quantum Theory, \href{http://arxiv.org/abs/arXiv:0708.3710}{arXiv:0708.3710} (2007)


\bibitem{MerminIthaca} Mermin, N. D.: The Ithaca Interpretation of Quantum Mechanics, Pramana 51, 549-565 (1998) \href{http://arxiv.org/abs/arXiv:quant-ph/9609013}{arXiv:quant-ph/9609013} 


\bibitem{Nelson} E. Nelson, Derivation of the Schr\"{o}dinger Equation from Newtonian Mechanics, Phys.Rev. 150, 1079-1085 (1966)



\bibitem{Saunders} Saunders, S.: Time, Decoherence and Quantum Mechanics. Synthese 102, 235-266 (1995)


\bibitem{clm} Schmelzer, I.: A Condensed Matter Interpretation of SM Fermions and Gauge Fields, Foundations of Physics, vol. 39, 1, p. 73, \href{http://arxiv.org/abs/arXiv:0908.0591}{arXiv:0908.0591} (2009) 



\bibitem{TegmarkMath} Tegmark, M.: The Mathematical Universe, Found Phys 38: 101-150 (2008) 


\bibitem{ValentiniNoDenial} Valentini, A.: De Broglie-Bohm Pilot-Wave Theory: Many Worlds in Denial?  in \cite{ManyWorlds2009}, \href{http://arxiv.org/abs/arXiv:0811.0810}{arXiv:0811.0810}

\bibitem{BrownDenial} Brown, H.R.: Comment on Valentini, ``De Broglie-Bohm Pilot-Wave Theory: Many Worlds in Denial?'',  in \cite{ManyWorlds2009}, \href{http://arxiv.org/abs/arXiv:0901.1278}{arXiv:0901.1278}

\bibitem{ManyWorlds2009} Saunders, S., Barrett, J., Kent, A., Wallace, D. (eds.), Many
Worlds? Realism, Everett, and quantum mechanics, Oxford University Press (2009)

\bibitem{Wallace} Wallace, D.: Worlds in the Everett interpretation, Studies in the History and Philosopy of Modern Physics 33, 637–661, \href{http://arxiv.org/abs/arXiv:quant-ph/0103092}{arXiv:quant-ph/0103092} (2002)

\bibitem{WallaceMeasurement} Wallace, D.: The quantum measurement problem: state of play,  \href{http://arxiv.org/abs/arXiv:0712.0149}{arXiv:0712.0149} (2007)

\bibitem{Weinberg} Weinberg, S.: Living in the Multiverse, \href{http://arxiv.org/abs/arXiv:hep-th/0511037v1}{arXiv:hep-th/0511037v1}

\bibitem{Zeh} H.D. Zeh, Why Bohm’s Quantum Theory?  \href{http://arxiv.org/abs/arXiv:quant-ph/9812059}{arXiv:quant-ph/9812059}




\bibitem{Exchange} An Exchange of Letters in Physics Today on ``Quantum Theory Without Observers'', February 1999, \href{http://www.math.rutgers.edu/~oldstein/papers/qtwoe/qtwoe.html}{www.math.rutgers.edu/$\sim$oldstein/papers/qtwoe/qtwoe.html}

\end{thebibliography}
\end{document}